\documentclass{article}
\usepackage[T1]{fontenc}
\usepackage[latin9]{inputenc}
\usepackage{geometry}
\usepackage{textcomp}
\usepackage{amstext}

\usepackage{graphicx}
\usepackage{color}
\usepackage[dvipdf]{epsfig}
\usepackage{amstext}
\usepackage{amsmath}
\usepackage{amssymb}

\begin{document}
\def\a{{\alpha }}
\def\g{{\gamma }}
\def\b{{\beta }}
\def\z{{\zeta }}
\def\zab{{\zeta_{ab}}}
\def\be{\begin{equation}}
\def\ee#1{\label{#1}\end{equation}}
\def\d{\textsf{d} }
\def\c{\textsf{c} }
\def\e{\textsf{e} }
\def\s{\textsf{s} }
\def\x{\textsf{x} }
 \def\bx{\mathbf{x} }
 \def\bp{\mathbf{p} }
 \def\p{\textsf{p} }
 \def\I{\textsf{I} }
  \def\n{\textsf{n} }
 \def\pp{\textsf{P} }
 \def\q{\textsf{q} }
\def\k{\textsf{k} }
 \def\J{\textsf{J} }
\def\no{\nonumber}
\def\lb{\label}
\def\h{\textsf{h} }
\def\x{\textsf{x} }
\def\D{\textsf{D}}
\def\lb{\label}
\def\Ki{{\rm Ki}}
\def\lab{{\langle\alpha\beta\rangle}}
\newcommand{\ben}{\begin{eqnarray}}
\newcommand{\een}{\end{eqnarray}}

\title{Transport coefficients for relativistic gas mixtures of hard-sphere particles }
\date{}

\author{ Gilberto M. Kremer%
\thanks{kremer@fisica.ufpr.br%
}, Valdemar Moratto%
\thanks{moratto.valdemar@gmail.com
}\\
 Departamento de F\'{\i}sica, Universidade Federal do Paran\'a, 81531-980  Curitiba,
Brazil}
\maketitle


\begin{abstract}
In the present work, we calculate the transport coefficients for a relativistic binary mixture of
diluted gases of hard-sphere particles. The gas mixture under consideration is studied within the
relativistic Boltzmann equation in the presence of a gravitational field described by the isotropic
Schwarzschild metric. We obtain the linear constitutive equations for the thermodynamic fluxes. The
driving forces for the fluxes of particles and heat will appear with terms proportional to the gradient
of gravitational potential. We discuss the consequences of the gravitational dependence on the driving
forces. We obtain general integral expressions for the transport coefficients and evaluate them by
assuming a hard-sphere interaction amongst the particles when they collide and not very disparate
masses and diameters of the particles of each species. The obtained results are expressed in terms
of their temperature dependence through the relativistic parameter which gives the ratio of the
rest energy of the particles and the thermal energy of the gas mixture. Plots are given to analyze
the behavior of the transport coefficients with respect to the temperature when small variations in
masses and diameters of the particles of the species are present. We also analyze for each coefficient
the corresponding limits to a single gas so the non-relativistic and ultra-relativistic limiting cases
are recovered as well. Furthermore, we show that the transport coefficients have a dependence on the
gravitational field.
\end{abstract}

 \section{Introduction}

An inspection of the literature (see e.g. the references \cite{1,St,VL,Her1,Her2,Gui,LK,Kox1,Kox2,And,GLW,CK})
concerning the analysis of relativistic gas mixtures shows that in the majority of the works,
only general expressions for the transport coefficients have been given.
One interesting feature of the transport coefficients of relativistic gases is their dependence
on a parameter $\z_a=m_a c^2/kT$ which gives the ratio of the rest energy of the particles of species $a$ and the thermal energy of the gas. This ratio is small for high temperatures so the gas is in the
ultra-relativistic regime, for low temperatures such a ratio is large and then the gas is in the
non-relativistic regime. For a single gas the expressions for the transport coefficients in
these limits are well known (see e.g. \cite{St,GLW,CK}). For mixtures of relativistic gases
there are only few works which analyze the transport coefficients. We quote: ref. \cite{KK}
where the reaction rate coefficient was obtained by considering a relativistic reactive differential
cross-section which takes into account the activation energy of the chemical reaction; ref. \cite{KM}
where Grad's moment method was employed for the determination of the transport coefficients for mixtures
of Maxwellian particles; lastly ref. \cite{KD} where the diffusion coefficient was determined by using
a BGK-type kinetic model for a mixture of hard-sphere  particles.

Some researches of relativistic gases in the presence of gravitational fields have been
recently performed, we quote for instance the works \cite{KD,AL,MS,KS,VG}. Among the results,
it is interesting to note that relativistic effects arise in the form of the thermal and diffusion
generalized forces. Such forces have a contribution due to the four-acceleration as originally
presented by Eckart \cite{Eck} for the thermal force and a gravitational potential gradient in
accordance with Tolman's law \cite{To1,To2}.

This report focuses the evaluation of the transport coefficients for a binary mixture of relativistic ideal gases and represents a continuation of the research presented in ref. \cite{VG}. In such a reference we used a method of solution for the Boltzmann equation that combines the Chapman-Enskog and Grad representations (see e.g. \cite{CK,GMK}). In ref. \cite{VG} we found the linear constitutive equations for the
heat and particle fluxes, dynamic pressure and pressure deviator tensor, and gave the transport coefficients as general integrals for a gas mixture. In the present paper we rewrite such expressions in
the particular case of a binary mixture and evaluate them by assuming two physical hypotheses:
1) a hard-sphere interaction of the particles when they collide and 2) not very disparate molecular masses and diameters for the species. The performance of the integrals for the determination of the transport coefficients represents long and tricky manipulations. For this reason we have added in the appendix A sufficient hints to reproduce all the calculations. In appendix B we have listed a number of integrals that appear along the process taking into account the two physical hypotheses mentioned before.
We present the expressions for the transport coefficients when they depend on differences of masses and diameters of the particles, and on concentrations of the species. We explore their behavior by analyzing some graphics with respect to the relativistic parameter $\z_a$. We analyze the one-species limits and we show
that they are in accordance with those reported in the literature \cite{CC,HCB,GMK}. Of course this last
statement guarantees the correct ultra-relativistic (high temperatures) and non-relativistic (low
temperatures) limits for each transport coefficient.
We also show that, due to the presence of the gravitational field, the transport coefficients become smaller.

Applications of these results can be done to some astrophysical situations like white dwarfs or clouds
nearby a source of gravitational potential. An example of self-diffusion can be addressed to a situation
in which different isotopes of the same gas diffuse through each other; in this case the mass difference
between the components of the gas can be very small.

The structure of the work is as follows. In section 2 we recall the basic equations and definitions. In
section 3 we analyze the shear and bulk viscosity coefficients for a binary mixture and the limiting cases for a simple fluid are given. Section 4 is devoted to the determination of the thermal conductivity, diffusion and thermal-diffusion rate coefficients; there the one-component limits are also given. The
influence of the gravitational field on the transport coefficients is presented in section 5 and finally the main conclusions  are stated in section 6.
\section{Background}

In this work we deal with a binary mixture of diluted ideal gases in the nearby of a gravitational potential
produced by a spherical static source. We assume a curved space-time described with the
isotropic Schwarzschild metric $g^{\mu\nu}$:
\ben\label{1}
ds^2=g_0(r)\left(dx^0\right)^2-g_1(r)\delta_{ij}dx^idx^j,\qquad
g_0(r)=\left(\frac{1+\frac{\Phi}{2c^2}}{1-\frac{\Phi}{2c^2}}\right)^2,\qquad
g_1(r)=\left(1-\frac{\Phi}{2c^2}\right)^4.
\een
The above equations contain the gravitational potential $\Phi=-GM/r$, where $G$ is the gravitational
constant, $M$ the total mass of the spherical source, $r$ the corresponding radius and $c$ the
light speed and indexes $\{i,j\}$ will run over spatial components.

The gases under consideration are composed by particles that do not have internal degrees of freedom and are characterized by their space-time coordinates $x^{\mu}=\left(ct , \textbf{x}\right)$. The four-momentum
of each particle is $p_a^{\mu}=\left(p_a^0 , \textbf{p}_a\right)$ where the Latin subindex $a=1,2$ denotes the species of the gas whereas the Greek index $\mu$ denotes the tensor properties. The momentum of each particle
 represents a time-like four-vector that holds the so-called mass-shell condition, i.e.,
 $g_{\mu\nu}p_a^\mu p_a^\nu =m_a^2 c^2$, where $m_a$ is the rest mass of a particle of species $a$.
 This last condition leads to a relation between the temporal and spatial components of the four-momentum
 in the following way:
\ben
p_{a}^{0}=p_{a0}/{g_{0}},\qquad\text{and}\qquad p_{a0}=\sqrt{g_{0}\left(m_{a}^{2}c^{2} -g_{1}
\vert{\bf p}_{a}\vert^{2}\right)}
\een
for the contravariant and covariant representations, respectively.

Statistical mechanics has as a basis the macroscopic description of a gas through a distribution function. As usual we will use the one-particle distribution function $f_{a}\left(x^{\mu},p_{a}^{\mu}\right)$
so that $f_{a}\left(x^{\mu},p_{a}^{\mu}\right){d}^{3}x\,{d}^{3}p_{a}$ is the number of particles of the
constituent $a$ in the volume element between $\bf x$, ${\bf x}+{d}^{3}x$ and
$\textbf{p}_{a}$, $\textbf{p}_{a}+{d}^{3}p_{a}$ at some instant of time $t$. This last quantity $f_a$
can be obtained by solving the Boltzmann equation, which for the case of a binary mixture in a
Riemannian space and in an appropriate manifestly covariant language reads \cite{GLW,CK}:
\ben\lb{be}
p_{a}^{\mu}\frac{\partial f_{a}}{\partial x^{\mu}}-\Gamma_{\mu\nu}^{i}p_{a}^{\mu}p_{a}^{\nu}
\frac{\partial f_{a}}{\partial p_{a}^{i}} =\sum_{b=1}^{2}\int(f_{a}'f_{b}'-f_{a}f_{b})F_{ba}
\sigma_{ab}\, d\Omega\sqrt{-g}\frac{d^{3}p_{b}}{p_{b0}}.
\een
Here we have introduced the Christoffel symbols $\Gamma_{\mu\nu}^{i}$. The invariant flux is
$F_{ba}=\sqrt{(p_{a}^{\mu}p_{b\mu})^{2}-m_{a}^{2}m_{b}^{2}c^{4}}$ and the invariant differential elastic cross section for collisions of species $a$ and $b$ is $\sigma_{ab}d\Omega$ where $d\Omega$
is the corresponding solid angle element. We have the invariant differential element
$\sqrt{-g}\frac{d^{3}p_{b}}{p_{b0}}$ with $\sqrt{-g}=\hbox{det}\left[g^{\mu\nu}\right]$.
Quantities denoted with a prime are evaluated with the momentum of the particles after a binary
collision occurs, that is, $f'_{a}\equiv f({\bf x},{\bf p}'_{a},t)$.

Following the standard processes \cite{GLW,CK} it is possible to obtain balance equations from
Boltzmann's equation and therein some basic definitions are made. Here we remind the general setup.
The four-flux of particles for the species $a$ is
\begin{equation}
N_{a}^{\mu}=c\int p_{a}^{\mu}f_{a}\sqrt{-g}\frac{d^{3}p_{a}}{p_{a0}},\qquad\hbox{and for the mixture}
\qquad N^{\mu}=\sum_{a=1}^{2}N_a^{\mu}.
\end{equation}
Here we shall use the Eckart frame \cite{Eck}, in which the hydrodynamic four-velocity  $U^\mu$ is
introduced with the following decomposition of
$N_{a}^{\mu}$ as
\begin{equation}
N_{a}^{\mu}=\n_{a}U^{\mu}+\J_{a}^{\mu},\qquad\hbox{where}\qquad \n_{a}=\frac{N_{a}^{\mu}U_{\mu}}{c^{2}}
\label{6}
\end{equation}
is the local number of particles of species $a$.
The definition of the diffusive particle four-flux $\J_{a}^{\mu}$ is
\begin{equation}
\J_{a}^{\mu}=\Delta_{\nu}^{\mu}c\int p_{a}^{\nu}f_{a}\frac{d^{3}p_{a}}{p_{a0}},\qquad\hbox{and it holds}
\qquad \J_{a}^{\mu}U_{\mu}=0.
\end{equation}
The projector has been introduced as
\ben\label{projetor}
\Delta^{\mu\nu}=g^{\mu\nu}-\frac{1}{c^{2}}U^{\mu}U^{\nu},\qquad\hbox{with the property}\qquad
\Delta^{\mu\nu}U_\mu=0.
\een
The diffusive particle four-flow is constrained by $\J_1^\mu+\J_2^\mu=0$, so that there exists
only one linearly independent diffusive particle four-flow for a binary mixture.

Other important definitions must be reminded. The energy-momentum tensor for one of the species
of the mixture is
\ben
T_a^{\mu\nu}=c\int p_{a}^{\mu}p_a^\nu f_{a}\sqrt{-g}\frac{d^{3}p_{a}}{p_{a0}},
\qquad\hbox{and for the mixture}\qquad T^{\mu\nu}=\sum_{a=1}^{2}T_a^{\mu\nu}.
\een
Following the literature, for instance ref. \cite{VG} and references therein,
the energy-momentum tensor for the mixture can be written as:
\ben\label{tem}
T^{\mu\nu}=\frac{\n\e}{c^{2}}U^{\mu}U^{\nu}
-(\p+\varpi)\Delta^{\mu\nu}+\p^{\langle\mu\nu\rangle} +\frac{1}{c^{2}}\left(U^{\mu}\q^{\nu}+U^{\nu}\q^{\mu}
\right),
\een
provided with definitions
\ben\lb{9}
\p^{\langle\mu\nu\rangle}=\sum_{a=1}^2\p_a^{\langle\mu\nu\rangle}
=\left(\Delta_{\sigma}^{\mu}\Delta_{\tau}^{\nu}
-\frac{1}{3}\Delta^{\mu\nu}\Delta_{\sigma\tau}\right)\sum_{a=1}^2 T_{a}^{\sigma\tau},
\qquad \q^{\mu}=\sum_{a=1}^2 \left(\q_a^{\mu}+\h_a\J_{a}^{\mu}\right)=\sum_{a=1}^2
\Delta_\sigma^\mu T_a^{\sigma\nu}U_\nu,
\\\lb{10}
\p+\varpi=\sum_{a=1}^2 (\p_{a}+\varpi_a)= -\frac13 \sum_{a=1}^2\Delta_{\mu\nu}
T_{a}^{\mu\nu},\qquad\qquad \n\e=\sum_{a=1}^{2}\n_{a}\e_{a}=\sum_{a=1}^{2}\frac{1}{c^2}U_\mu
T_a^{\mu\nu}U_\nu.
\een
Here we have introduced: the partial pressure deviator tensor $\p_a^{\langle\mu\nu\rangle}$,
the partial heat flux $\q_a^\mu$, the partial enthalpy $\h_a=\e_a+\p_a/\n_a$, the internal energy per particle for the mixture $\n\e=\n_1\e_1+\n_2\e_2$, the local partial pressure $\p_a=\n_a kT$ and the partial dynamic pressure $\varpi_a$. Of course quantities without Latin subindex refer to the mixture.

Equations (\ref{9}) and (\ref{10}) depend on the form of the solution of the Boltzmann equation
for the distribution function
$f_a$ and in this work we have used a linear approximation that have been obtained in \cite{VG}
with the thermodynamic local variables $\{\n_1,\n_2,U^\mu,T\}$, being $T$ the local temperature.
Details of such approximation to the solution can be found in the cited reference.
Let us mention here, taking as a basis the results of ref. \cite{VG}, the process to obtain
the constitutive equations for the thermodynamical fluxes.

Firstly, to obtain the Fick and Fourier laws we address Eqs. (48) and (55) from ref. \cite{VG}.
Such equations, when applied to a binary mixture and after some algebraic manipulations
lead us to the following linear system for the heat and particle partial fluxes:

\ben
-\frac{\n_1+\n_2}{\n_1 \n_2}\d^\mu_1\p=\left(\mathcal{A}_{11}-2\mathcal{A}_{12}
+\mathcal{A}_{22}\right)\J_{1}^{\mu}
-\left(\mathcal{F}_{11}-\mathcal{F}_{21}\right)\q_{1}^{\mu}-\left(\mathcal{F}_{12}
-\mathcal{F}_{22}\right)\q_{2}^{\mu},
\label{11}
\\
\frac{\nabla^{\mu}\mathcal{T}}{T}=\left(\mathcal{F}_{11}-\mathcal{F}_{21}\right)\J_{1}^{\mu}
-\mathcal{H}_{11}\q_{1}^{\mu}-\mathcal{H}_{12}\q_{2}^{\mu}\label{12},
\\
\frac{\nabla^{\mu}\mathcal{T}}{T}
=\left(\mathcal{F}_{12}-\mathcal{F}_{22}\right)\J_{1}^{\mu}-\mathcal{H}_{21}\q_{1}^{\mu}
-\mathcal{H}_{22}\q_{2}^{\mu}.\label{13}
\een
Here we remind that number indexes refer to a species of the mixture, the total pressure
is $\p=(\n_1+\n_2)kT$. The thermodynamic force directly related with diffusion arises as
\ben
\d_1^{\mu}=\nabla^\mu\x_{1}+\left(\x_{1}-1\right)\nabla^\mu\ln\p
-\frac{\n_{1}\h_{1}-\n\h}{\p c^{2}}\Delta^{\mu j}\left[U^{\tau}
\frac{\partial U_{j}}{\partial x^{\tau}}-\frac{1}{1-{\Phi^2}/4c^{4}}
\frac{\partial{\Phi}}{\partial x^{j}}\right],\label{df}
\een
where $\x_1=\p_1/\p=\n_1/\n$ is the concentration of species labeled by 1
and $\nabla^\mu=\Delta^{\mu\nu}\partial_\nu$ is the gradient operator. Equations (\ref{12})
and (\ref{13}) refer to a generalized thermal force defined as
\ben
\nabla^{\mu}\mathcal{T}=\nabla^\mu T-\frac{T}{c^{2}}\Delta^{\mu i}\left[U^{\nu}
\frac{\partial U_{i}}{\partial x^{\nu}}-\frac{1}{1-{\Phi^2}/4c^{4}}
\frac{\partial{\Phi}}{\partial x^{i}}\right].\label{tf}
\een
As mentioned in the introduction, the generalized diffusive and thermal forces (\ref{df}) and (\ref{tf}) contain terms of relativistic nature as well as gravitational contributions. There are some
comments to be made about the significance and physical consequences of these new terms and
we are leaving them to the conclusion section.

Secondly, a process described in \cite{VG} leads to the obtention of the constitutive equation
for the viscosities. The bulk viscosity for a binary mixture emerges through
\ben\lb{16}
-\left[\frac{\p_1 kT}{c^3}\frac{\partial\ln\c_v^1}{\partial \ln\z_1}\right]\nabla_\mu U^\mu=
\mathcal{R}_{11}\varpi_1+\mathcal{R}_{12}\varpi_2,\qquad -\left[\frac{\p_2 kT}{c^3}
\frac{\partial\ln\c_v^2}{\partial \ln\z_2}\right]\nabla_\mu U^\mu=\mathcal{R}_{21}\varpi_1
+\mathcal{R}_{22}\varpi_2.
\een
Above we have introduced the specific heat at constant volume
$\c_{v}^{a}=k\left(\zeta_{a}^2+5G_{a}\z_a-G_{a}^{2}\zeta_{a}^2-1\right)$ of species $a=1,2$
with $G_a={K}_3(\z_a)/{K}_2(\z_a)$. ${K}_n$ are modified Bessel functions
of the second kind and order $n$ (see definition (\ref{bf}) of Appendix A).

The shear viscosity emerges from the definition of the partial pressure deviator tensors
$\p^{\langle\mu\nu\rangle}_1$:
\ben\lb{17}
2\nabla^{\langle\mu}U^{\nu\rangle} =\mathcal{K}_{11}\p_1^{\langle\mu\nu\rangle}
+\mathcal{K}_{12}\p_2^{\langle\mu\nu\rangle},
\qquad
2\nabla^{\langle\mu}U^{\nu\rangle} =\mathcal{K}_{21}\p_1^{\langle\mu\nu\rangle}
+\mathcal{K}_{22}\p_2^{\langle\mu\nu\rangle}.
\een
The set (\ref{11}) - (\ref{13}), (\ref{16}) and (\ref{17}) constitutes an algebraic system for the
constitutive equations. The functions $\mathcal{A}'s$, $\mathcal{F}'s$, $\mathcal{H}'s$, $\mathcal{R}'s$
and $\mathcal{K}'s$ are general integrals that depend on the interaction of the particles when they collide.
 We have listed them in the appendix B where the first equality represents the general case whereas
 in the second one they are  evaluated for the case of hard-sphere interaction and non disparate
 masses and diameters.

\section{Navier-Stokes law}

In this section we analyze the constitutive equations for the pressure deviator tensor
and dynamical pressure for a diluted binary mixture. Such equations emerge by solving the
two algebraic systems represented in Eqs. (\ref{16}) and (\ref{17}) and taking the sum over the species.
We obtain
\ben\label{nsce}
\p^{\langle\mu\nu\rangle}=2\mu\nabla^{\langle\mu} U^{\nu\rangle}\qquad\text{and}\qquad\varpi
=-\eta\nabla_\mu U^\mu,
\een
where the total shear and bulk viscosities are $\mu$ and $\eta$, respectively. The transport coefficients
read
\ben\label{ns2a}
\mu=\frac{\mathcal{K}_{11}-2\mathcal{K}_{12}+\mathcal{K}_{22}}{\mathcal{K}_{11}\mathcal{K}_{22}
-\mathcal{K}_{12}^2},
\\\label{ns2b}
\eta=\frac{kT\left(\mathcal{R}_{22}-\mathcal{R}_{21}\right)\p_1}
{c^3(\mathcal{R}_{11}\mathcal{R}_{22}-\mathcal{R}_{12}\mathcal{R}_{21})}
\left[\frac{\partial\ln \c_v^1}{\partial\ln\z_1}
+\frac{\left(\mathcal{R}_{11}-\mathcal{R}_{12}\right)\n_2}{\left(\mathcal{R}_{22}-\mathcal{R}_{21}\right)
\n_1}\frac{\partial\ln \c_v^2}{\partial\ln\z_2}\right],
\een
where the elements of the matrices $\mathcal{K}_{ab}$ and $\mathcal{R}_{ab}$ are integrals given in
Appendix B (See Eqs. (\ref{A20h}) -- (\ref{A20m})). Furthermore, we have introduced the derivative
of the heat capacity per particle at constant volume:
\be
\frac{\partial\ln \c_v^a}{\partial\ln\z_a}=\frac{\z_a(20G_a+3\z_a
-13G_a^2\z_a-2G_a\z_a^2+2G_a^3\z_a^2)}{(1+G_a^2\z_a^2-\z_a^2-5G_a\z_a)},\qquad a=1,2.
\ee{ns2c}

If we consider that  the rest masses, the particle number densities and  the differential cross-sections
of both constituents are the same  -- i.e., $m_1=m_2=m$, $\n_1=\n_2=\n$
and $\sigma_{11}=\sigma_{22}=\sigma_{12}=\sigma$ -- we get  that the shear (\ref{ns2a})
and bulk (\ref{ns2b}) viscosities reduce to their expressions of a single constituent,
namely \cite{GLW,CK}
\ben\lb{ns3a}
\mu=\frac{15}{64\pi}\frac{kT}{c\sigma}\frac{\z^4K_3(\z)^2}{(2+15\z^2)K_2(2\z)+(3\z^3+49\z)K_3(2\z)},
\\\lb{ns3b}
\eta=\frac1{64\pi}\frac{kT}{c\sigma}\frac{\z^4K_2(\z)^2}{2K_2(2\z)+\z K_3(2\z)}
\frac{(20G+3\z-13G^2\z-2G\z^2+2G^3\z^2)^2}{(1-5G\z-\z^2+G^2\z^2)^2}.
\een

The limiting cases of low temperature (non-relativistic $\z\gg1$) and high temperature
(ultra-relativistic $\z\ll1$) are well known and can be obtained from (\ref{ns3a}) and (\ref{ns3b})
as it is shown in the corresponding literature \cite{CK}.

\begin{figure}
\centering
\vskip0.5cm
\includegraphics[width=13cm]{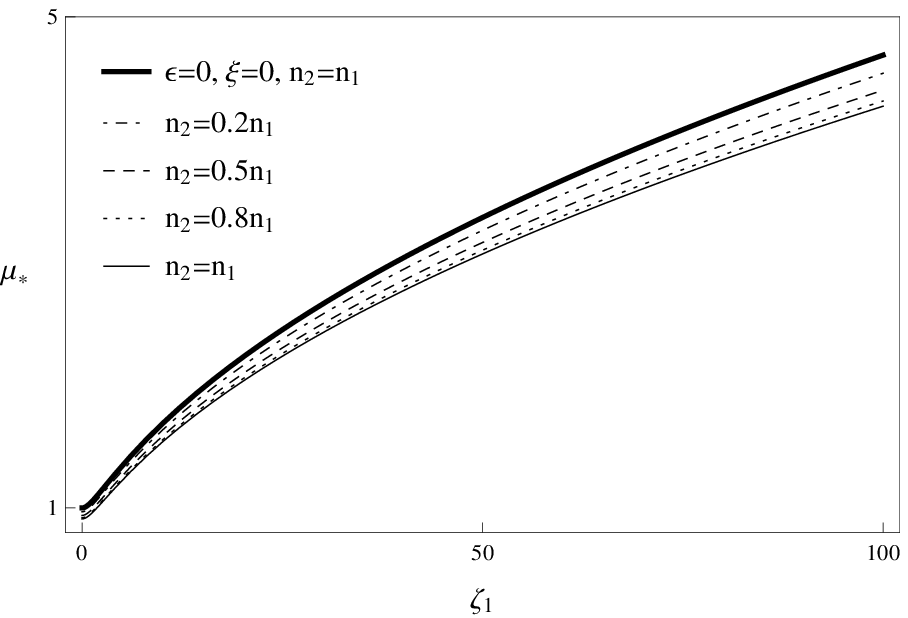}
\caption{Dimensionless shear viscosity $\mu_*$  as function of the parameter $\z_1=m_1c^2/kT$.
The thick line represents the single fluid and the others are for the  mixture with
$\epsilon=0.01$, $\xi=0.1$ and different ratios $\n_2/\n_1$.}
\lb{fig1a}
\end{figure}

\begin{figure}
\centering
\vskip0.5cm
\includegraphics[width=13cm]{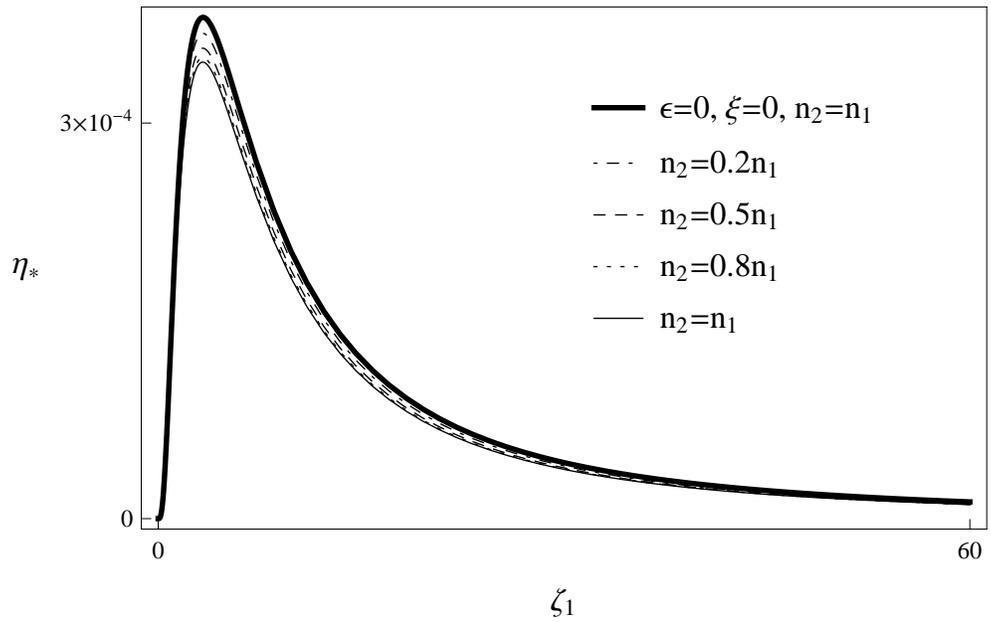}
\caption{Dimensionless bulk viscosity $\eta_*$ as function of the parameter $\z_1=m_1c^2/kT$. The thick line
 represents the single fluid and the others for the mixture with $\epsilon=0.01$, $\xi=0.1$ and
 different ratios $\n_2/\n_1$.}
\lb{fig1b}
\end{figure}
To determine the $\mathcal{K}'s$ and $\mathcal{R}'s$ functions contained in the the transport
coefficients ((\ref{ns2a}) and (\ref{ns2b})) we  incorporate two physical hypotheses.
Firstly we consider similar masses of the particles of different species, so we can write
$m_2=m_1(1+\epsilon)$ with $\epsilon\ll1$. Secondly we  assume hard-spheres,
so the diameters of the particles are constant. Therefore a small difference is assumed for
the diameter of a particle of the species 2 with respect to the diameter of a particle of the species 1.
That is $\d_2=\d_1(1+\xi)$ with $\xi\ll1$. The hard-sphere differential cross sections can be written
according to $\sigma_{11}=\d_1^2/4$, $\sigma_{22}=\d_2^2/4$ and $\sigma_{12}=(\d_1+\d_2)^2/16$.
Consequently we have the relations $\sigma_{11}=\sigma$, $\sigma_{12}=\sigma(1+\xi)$ and
$\sigma_{22}=\sigma(1+2\xi)$ with $\sigma$=constant. The evaluation of the integrals is a long task,
we have depicted the principal steps in the appendix A whereas the second identity of the list
of appendix B represents the expressions finally evaluated.

By substituting the corresponding functions of appendix B into the shear (\ref{ns2a}) and bulk
(\ref{ns2b}) viscosities and rewriting them as dimensionless quantities in the form
$\mu_*=10\mu\,c\sigma_{11}/3kT$ and $\eta_*=\eta\,c\sigma_{11}/kT$, we can  plot them with respect
to the relativistic parameter $\z_1$. Figure \ref{fig1a} shows the behavior of $\mu_*$ as function
of $\z_1$ when $\epsilon=0.01$, $\xi=0.1$ and different concentrations of the ratio $\n_2/\n_1$ are imposed.
 On the other hand we  plot the dimensionless bulk viscosity in Figure \ref{fig1b} where the same
 conditions for the mass and concentration differences were assumed.  We shall  leave
the comments of the figures to the conclusion section. Now we shall focus on the thermal
conductivity, diffusion and thermal diffusion.

\section{Fourier and Fick laws}

In this section we analyze the constitutive equations for the heat and
diffusive fluxes and the corresponding transport coefficients. We shall follow a similar
process of the last section in which the general expressions are given, the one-component
limit is recovered and a plot of the general expressions is shown. In the present article
we are interested in the description of the Fick and Fourier laws. The Fick law describes the
flux of particles due to the thermal and diffusive thermodynamic forces, see Eqs. (\ref{df}) and (\ref{tf}).
The Fourier law establishes that the thermal conductivity is the proportionality coefficient between the
heat flux and the thermal force in the absence of diffusion. Given the objective of the present
section we start by solving the system of equations (\ref{11})-(\ref{13}) for the heat flux and
neglecting the diffusion terms, then we obtain the Fourier law for the case of a binary mixture:
\ben\label{ff3}
\q^\mu=\lambda\nabla^\mu\mathcal{T},\qquad \hbox{where}\qquad
\lambda=-\frac{\mathcal{H}_{11}+\mathcal{H}_{22}-2\mathcal{H}_{12}}
{T(\mathcal{H}_{11}\mathcal{H}_{22}-\mathcal{H}_{12}^2)},
\een
here $\lambda$ is the thermal conductivity. The $\mathcal{H}'s$ functions are given by Eqs.
(\ref{A20e}) -- (\ref{A20g}) in the Appendix B.
To analyze the limit to a single component we again consider the conditions $m_1=m_2=m$,
$\n_1=\n_2=\n$ and $\sigma_{11}=\sigma_{22}=\sigma_{12}=\sigma$. Then the thermal conductivity
coefficient reduces to
\ben\lb{ff4}
\lambda=\frac3{64\pi}\frac{ck}\sigma\frac{(\z+5G-G^2\z^2)^2\z^4K_2(\z)^2}{(\z^2+2)K_2(2\z)+5\z K_3(2\z)},
\een
which is in accordance with the expression \cite{GLW,CK} obtained when the kinetic theory is applied
to a single fluid. Of course this last expression guarantees the corresponding limiting cases for
low temperatures (non-relativistic $\z\gg1$) and high temperatures (ultra-relativistic $\z\ll1$).

Figure \ref{fig2} shows the dimensionless thermal conductivity $\lambda_*=\lambda\,c\sigma_{11}m_1/k^2T$
as function of $\z_1$ with $\epsilon=0.01$, $\xi=0.1$  and different ratios $\n_2/\n_1$.
Some comments about this graphic are given in the conclusion section.

\begin{figure}
\centering
\vskip0.5cm
\includegraphics[width=13cm]{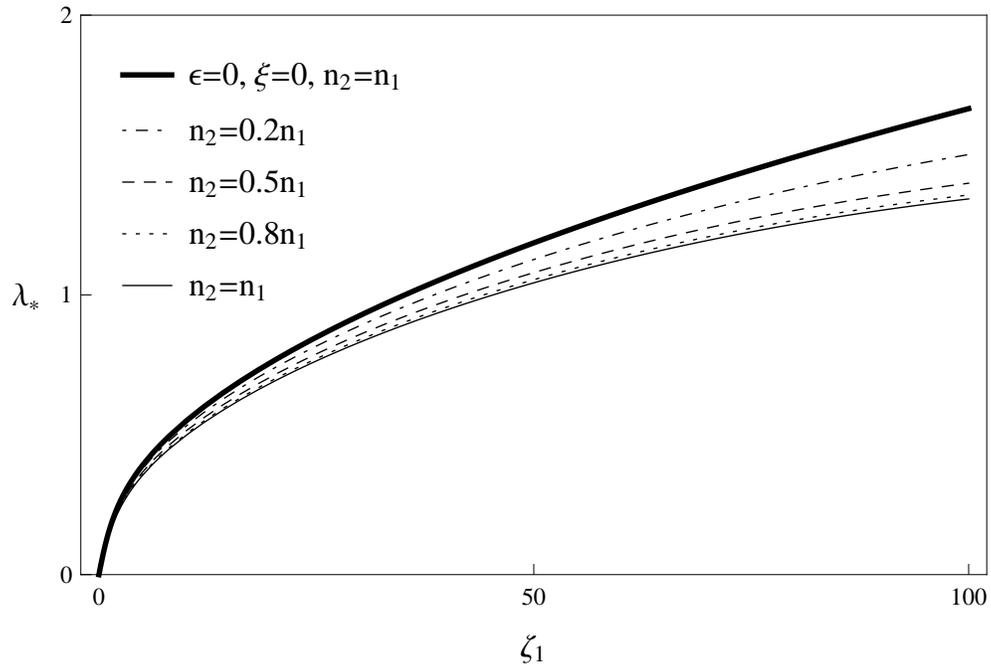}
\caption{Dimensionless thermal conductivity $\lambda_*$
as function of the parameter $\z_1=m_1c^2/kT$. The thick line represents the single fluid.
Different scenarios for the mixture are plotted with $\epsilon=0.01$, $\xi=0.1$ and some ratios $\n_2/\n_1$. }
\lb{fig2}
\end{figure}

Now we focus on the Fick law.  For the case of the binary mixture we have only one diffusion flux
as it can be appreciated from Eqs. (\ref{11})-(\ref{13}). By solving for the particle flux we obtain:
\ben\label{ff5}
\J^\mu=\n D\left(\d^\mu+\frac{k_T}T\nabla^\mu\mathcal{T}\right),
\een
where we have redefined $\J_1^\mu$ as $\J^\mu$ and $\d_1^\mu$ as $\d^\mu$. Here $D$ and $k_T$
are the coefficients of diffusion and thermal-diffusion ratio, respectively. They are given by
\ben\label{ff6a}
D=\frac{kT/\n}{\mathcal{A}_{12}+\frac{\mathcal{F}_{12}\mathcal{F}_{21}}{\mathcal{H}_{11}\mathcal{H}_{22}
-\mathcal{H}_{12}^2}\left[\frac{\n_2\mathcal{F}_{21}}{\n_1\mathcal{F}_{12}}\mathcal{H}_{22}
+\frac{\n_1\mathcal{F}_{12}}{\n_2\mathcal{F}_{21}}\mathcal{H}_{11}+2\mathcal{H}_{12}
\right]},
\een
\ben\label{ff6b}
k_T=\frac1{\p}\left[\frac{\n_2\mathcal{F}_{21}(\mathcal{H}_{12}
-\mathcal{H}_{22})+\n_1\mathcal{F}_{12}(\mathcal{H}_{11}
-\mathcal{H}_{12})}{\mathcal{H}_{11}\mathcal{H}_{22}-\mathcal{H}_{12}^2}\right],
\een
where the elements of the matrices $\mathcal{A}_{ab}, \mathcal{F}_{ab}, \mathcal{H}_{ab}$
are given by the expressions (\ref{A20a}) -- (\ref{A20g}) of the Appendix B.

Now we can obtain the one-species limit from the general Fick law Eq. (\ref{ff5}) by setting
$m_1=m_2=m$, $\n_1=\n_2=\n$, and $\sigma_{11}=\sigma_{22}=\sigma_{12}=\sigma$. Equation (\ref{ff6a})
leads to the self-diffusion coefficient, but it is quite long to be shown here.
Its low temperature (non-relativistic $\z\gg1$) and high temperature (ultra-relativistic $\z\ll1$)
limiting cases read
\ben\lb{ff7a}
D=\frac3{8\n\d^2}\sqrt{\frac{kT}{\pi m_1}}\left(1-\frac8{15\z}+\dots\right),\qquad \z\gg1,
\\\lb{ff7b}
D=\frac{6c}{13\pi\n\d^2}\left(1-\frac{2\z^2}{13}+\dots\right),\qquad \z\ll1.
\een
Note that the low temperature self-diffusion coefficient (\ref{ff7a}) is proportional to
$\sqrt{T}/\n$ in accordance with the first non-relativistic approximation for a gas of
hard-sphere particles \cite{CC,HCB,GMK}. On the other hand, the one-species limit for the
thermal-diffusion coefficient $k_T$ goes to zero (note that $\mathcal{H}_{11}=\mathcal{H}_{22}$ and
$\mathcal{F}_{12}=\mathcal{F}_{21}$), as expected from the kinetic theory when applied to a
single component.

Figures \ref{fig3a} and \ref{fig3b} show the behavior of the dimensionless diffusion coefficient
$D_*=m_1D\,c\sigma\n/kT$ and the thermal-diffusion ratio $k_T$ as functions of the relativistic
parameter $\z_1$ in similar conditions for $\epsilon$ and $\xi$ as the previous plots.
Comments are left to the conclusions. In the following section we are going to show
the explicit dependence of the transport coefficients on the gravitational potential.

\begin{figure}
\centering
\vskip0.5cm
\includegraphics[width=13cm]{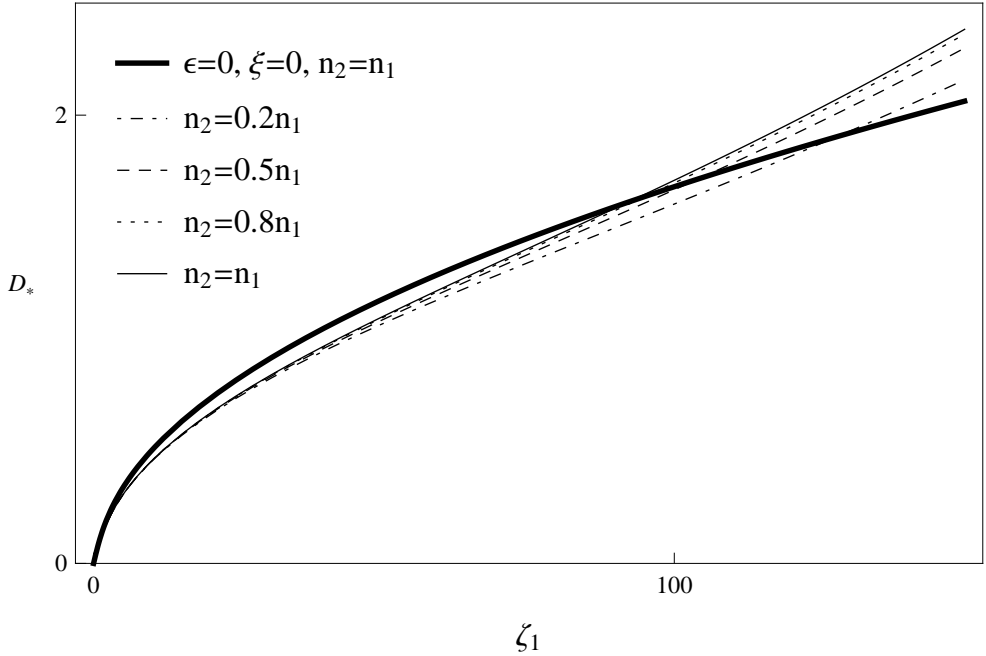}
\caption{Dimensionless diffusion coefficient $D_*$ as function of the parameter $\z_1=m_1c^2/kT$ for
$\epsilon=0.01$, $\xi=0.1$ and different ratios $\n_2/\n_1$.
The thick line represents the self-diffusion coefficient.}
\lb{fig3a}
\end{figure}

\begin{figure}
\centering
\vskip0.5cm
\includegraphics[width=13cm]{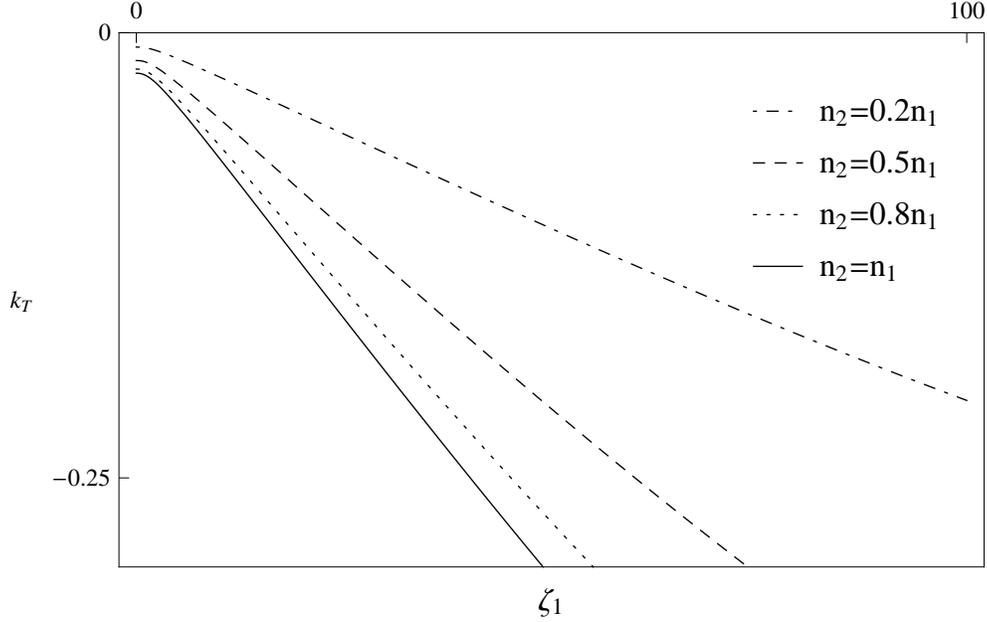}
\caption{Thermal-diffusion ratio $k_T$ as function of the parameter $\z_1=m_1c^2/kT$ for
$\epsilon=0.01$, $\xi=0.1$ and different ratios $\n_2/\n_1$.}
\lb{fig3b}
\end{figure}

\section{The gravitational field dependence on the transport coefficients}

In this section we explore the dependence of the transport coefficients on the gravitational field.
We quote two works \cite{KD,KS} where such a dependence arises naturally from the kinetic theory
point of view. Then more discussion is needed and here we analyze the corresponding transport
coefficients for a binary mixture.

Let us begin by noticing that the first term of the energy momentum tensor Eq. (\ref{tem})
can be written in the comoving frame, i.e. $U^\mu=(c/\sqrt{g_0},\bf0)$, as:
\ben\label{ne}
\frac{\n\e}{c^{2}}U^{\mu}U^{\nu} =\n\e\left(\frac{1-\frac{\Phi}{2c^2}} {1+\frac{\Phi}{2c^2}}\right)^2
\equiv \widetilde{\n\e},
\een
where the quantity $\widetilde{\n\e}$ is now playing the role of the internal energy density.

We can also write the projector $\Delta^{\mu\nu}=g^{\mu\nu}-U^\mu U^\nu/c^2$ with the
Schwarzschild metric (\ref{1}) in the comoving frame, in this case the components of the projector become
\ben\label{d1}
\Delta^{00}=0,\qquad \Delta^{ij}=-\frac{\delta^{ij}} {\left(1-\frac{\Phi}{2c^2}\right)^4}.
\een

The pressure tensor $\mathcal{P}^{\mu\nu}$ is defined as the second and third term of the energy-momentum
tensor Eq. (\ref{tem}). When the constitutive equations for the bulk and shear viscosities
(Eqs. (\ref{nsce})) are substituted in $\mathcal{P}^{\mu\nu}$ we have
\ben\label{tp}
\mathcal{P}^{\mu\nu}=-\p\Delta^{\mu\nu}+\eta \nabla_\gamma U^\gamma \Delta^{\mu\nu}+2\mu
\left( \frac{\Delta^\mu_\sigma\Delta^\nu_\tau+\Delta^\nu_\sigma\Delta^\mu_\tau}2
-\frac{\Delta^{\mu\nu}\Delta_{\sigma\tau}}3\right)\partial^\sigma U^\tau.
\een
The evaluation of Eq. (\ref{tp}) as well as the Fourier (\ref{ff3}) and Fick (\ref{ff5}) laws in the
comoving frame (\ref{d1}) and in Cartesian coordinates leads to
\ben\label{d2a}
&&\mathcal{P}^{ij}=\left[\widetilde\p
-\widetilde\eta\frac{\partial U^k}{\partial x^k}\right]\delta^{ij}
-\widetilde\mu\left[\frac{\partial U^i}{\partial x_j}+\frac{\partial U^j}{\partial x_i}
-\frac23\frac{\partial U^k}{\partial x^k}\delta^{ij}\right],
\\\label{d2b}
&&\q^i=-\widetilde\lambda\frac{\partial \mathcal{T}}{\partial x_i},
\qquad\qquad
\J^i=-\n\widetilde D\left[\d^i+\frac{k_T}T\frac{\partial \mathcal{T}}{\partial x_i}\right].
\een
Here the generalized thermal and diffusion forces read
\ben\lb{d3a}
\frac{\partial \mathcal{T}}{\partial x_i}&=&\frac{\partial T}{\partial x_i}-\frac{T}{c^2}\left(\dot U^i-\frac1{1-\Phi^2/4c^4}\frac{\partial \Phi}{\partial x_i}\right),
\\
\d^i&=&\frac{\partial \x_{1}}{\partial x_i}+\left(\x_{1}-1\right)\frac{\partial\ln\p}{\partial x_i}
+\frac{\n_{2}\h_{2}}{\p c^{2}}\left(\dot U^i-\frac1{1-\Phi^2/4c^4}\frac{\partial \Phi}{\partial x_i}\right),
\een
respectively. In the above equations $\dot U^i$ is the acceleration.

The coefficients with tildes in Eqs. (\ref{d2a}), (\ref{d2b}a) and (\ref{d2b}b) are defined as follows:
\ben\label{d4a}
\widetilde\p=\frac\p{\left(1-\frac{\Phi}{2c^2}\right)^4},
\qquad\widetilde\eta=\frac{\eta}{\left(1-\frac{\Phi}{2c^2}\right)^8},\qquad
\widetilde\mu=\frac\mu{\left(1-\frac{\Phi}{2c^2}\right)^8},
\\\label{d4b}
\widetilde\lambda=\frac\lambda{\left(1-\frac{\Phi}{2c^2}\right)^4},\qquad
\widetilde D=\frac{D}{\left(1-\frac{\Phi}{2c^2}\right)^4}.
\een
The above equations show a gravitational field dependence of the transport coefficients,
we leave the analysis to the conclusion section.

\section{Conclusions}

In this work we have evaluated analytically the transport coefficients for a relativistic binary mixture
of ideal gases in the presence of a gravitational field. This work represents the analysis of a particular
case (binary mixture) of the $r$-mixture studied by the authors in ref. \cite{VG} in which, among other
results, the transport coefficients were given in terms of general integrals. We applied the combined
Chapman-Enskog linear method of solution to the Boltzmann equation and obtain the transport coefficients
associated with the Navier-Stokes, Fourier and Fick laws. Such expressions are represented by general
integrals for the shear (\ref{ns2a}) and bulk (\ref{ns2b}) viscosities, thermal conductivity (\ref{ff3}b),
diffusion (\ref{ff6a}) and thermal-diffusion ratio (\ref{ff6b}). The process for the evaluation of the
integrals represents a hard task and in the appendix A such calculations are addressed. In the development of the
integrals we assumed two physical hypotheses: Firstly, the masses of the particles of each species are
similar i.e. $m_2=m_1(1+\epsilon)$ with $\epsilon\ll1$. Secondly, for the particles with hard-sphere
interaction,  we can assume that the diameters of the particles of each
species are very similar, i.e. $\d_2=\d_1(1+\xi)$ with $\xi\ll1$.
Once upon  the transport coefficients were evaluated, we presented a figure for each
one showing its dependence with respect to the relativistic parameter $\z_1$. Lastly we
analyzed  dependence of the transport coefficients on the gravitational potential and the corrections
for the pressure and internal energy. Now we shall comment the results.

Following the theory of fluid mixtures (see e.g. \cite{CC,HCB}) it is usual to express the diffusion
force in terms of gradients of pressure and concentration. From the expression for the generalized diffusive force obtained in this article, see Eq. (\ref{df}), we can identify four contributions to it:
1) a concentration gradient, 2) a pressure gradient, 3) terms proportional to the four-acceleration, and 4) terms proportional the gravitational potential gradient. The flux of particles or Fick's law is proportional to the diffusive and thermal forces (see Eq. (\ref{ff5})). Then we can conclude that the nature of
this transport is a consequence of:
\begin{itemize}
  \item A concentration gradient that tends to reduce the non-homogeneity of the mixture.
  \item A pressure gradient, where heavy particles tend to diffuse to places with high pressures,
  e.g. in centrifuges.
  \item A temperature gradient, that transports the matter to a warmer regions.
  \item An acceleration which acts on different masses.
  \item A gravitational potential gradient.
\end{itemize}
There is a very interesting physical situation that we can underline. In the particular case where the
acceleration is absent and the pressure and temperature are constant, the diffusion vanishes if the
gradient of concentration counterbalances the gradient of gravitational potential. On the other hand,
it is important to note that the last two terms of the diffusive force (\ref{df})  are a combination of
the acceleration and gravitational potential and are not of relativistic nature, since
\ben
\frac{(\n_a\h_a-\n\h)}{c^2\p}\rightarrow\frac{(\n_am_a-\sum_{b=1}^2\n_bm_b)}{\p}
\een
in the non-relativistic limiting case. Therefore we conclude that equation (\ref{df}) is
the generalization of the diffusion force originally written for the non-relativistic case \cite{CC,HCB}.

Let us focus now on the generalized thermal force Eq. (\ref{tf}). There we identify that it has three
contributions: 1) a pure gradient of temperature, 2) terms proportional to the four-acceleration and 3) terms proportional to the gravitational potential gradient. When the non-relativistic limit is analyzed, the term
that is a combination of the acceleration and gravitational potential goes to zero (contrary to the
case of the diffusive force) since it has a factor $T/c^2$ of relativistic order.
When the generalized thermal force (\ref{tf}) is substituted into the Fourier law (\ref{ff3}),
 an acceleration term $U^{\nu}\partial U_{i}/\partial x^{\nu}$ appears that coincides
with the one presented in a phenomenological analysis made by Eckart \cite{Eck}. Such a term
represents an isothermal heat flux when matter is accelerated.

Let us now analyze a consequence of the gravitational potential gradient
${\partial{\Phi}}/\partial x^{i}$ dependence on Eq. (\ref{tf}). First we analyze the
corresponding factor to such gradient in the comoving frame ($U^\mu=\left(c/\sqrt{g_0},\bf0\right)$),
by expanding it with $\Phi/c^2\ll1$ (weak field), so we have
\ben\label{ex}
\frac{\Delta^{ij}}{1-{\Phi^2}/4c^{4}}\simeq-\delta^{ij}\left[ 1+\frac{2\Phi}{c^2}
+\mathcal{O}\left(\frac{\Phi}{c^2}\right)^2\right].
\een
The Tolman law \cite{To1,To2} comes from phenomenological arguments, and states that at equilibrium, a gradient
of temperature is counterbalanced by a gravitational potential gradient. In a physical situation where
the heat flux and the acceleration term vanish we get from equation (\ref{tf}) when (\ref{ex})
is inserted at lowest order
\ben\label{ToL}
\frac{\nabla T}{T}=-\frac{\nabla\Phi}{c^2},
\een
i.e. we recover Tolman's law.
It is important to mention that, from the kinetic theory point of view, this fact was initially
presented in ref. \cite{AL,KD,KS,VG}.

Now let us pay attention on the thermodynamic variables. Note that the quantity
$\widetilde{\p}$ from equation (\ref{d4a}a) is playing the role of the pressure and it has
a factor that depends on the gravitational potential. If we expand for a weak field, i.e.
$\Phi/c^2\ll1$, we obtain
\ben\label{ptil}
\widetilde\p=\p\left(1+\frac{2\Phi}{c^2} +\mathcal{O}\left(\frac{\Phi}{c^2}\right)^2\right).
\een
In the same fashion, for $\Phi/c^2\ll1$, we expand Eq. (\ref{ne}) and obtain
\ben
\widetilde{\n\e}=\n\e\left(1-\frac{2\Phi}{c^2} +\mathcal{O}\left(\frac{\Phi}{c^2}\right)^2\right).
\een
Here at first order we have recovered the first post Newtonian approximation (1PN) for the pressure
and energy density as given by  Weinberg \cite{Wein}. Recently, it has been shown the same dependence
for the pressure by using the 1PN formalism in a different research \cite{KRW}.

On the other hand, we can analyze the gravitational dependence of the one-component limit of Eqs.
(\ref{d4a}b), (\ref{d4a}c), (\ref{d4b}a) and (\ref{d4b}b) for the bulk and shear viscosities,
thermal conductivity and diffusion, respectively. We expand for a low temperature (non-relativistic,
$\z\gg1$) and weak field $\Phi/c^2\ll1$ case and we get:
\ben\label{tcexp1}
\widetilde\eta&=&\frac{25}{64\d^2\z^2}\sqrt{\frac{mkT}\pi}\left(1- \frac{183}{16}\frac{1}{\z^2}+\cdots\right) \left(1-\frac{4GM}{c^2 r}+\frac{9(GM)^2}{c^4 r^2}-\cdots\right)\\\label{tcexp2}
\widetilde\mu&=&\frac5{16\d^2}\sqrt{\frac{mkT}\pi} \left(1+\frac{25}{16\z}+\dots\right)\left(1-\frac{4GM}{c^2 r}+\frac{9(GM)^2}{c^4 r^2}-\cdots\right)\\\label{tcexp3}
\widetilde\lambda&=&\frac{75}{64\d^2}\frac{k}{m}\sqrt{\frac{mkT}\pi} \left(1+\frac{13}{16\z}+\dots\right)\left(1-\frac{2GM}{c^2 r}+\frac{5(GM)^2}{2c^4 r^2}-\cdots\right)\\\label{tcex4}
\widetilde D&=&\frac{3}{8\n\d^2}\sqrt{\frac{kT}{\pi m_1}} \left(1+\frac{8}{15\z}+\dots\right)\left(1-\frac{2GM}{c^2 r}+\frac{5(GM)^2}{2c^4 r^2}-\cdots\right).
\een
Here we conclude that the transport coefficients become smaller in the presence of a gravitational field.
The  dependence of the thermal conductivity on the gravitational field has been studied and reported in ref. \cite{SG} from a
non-relativistic approach based on the Boltzmann equation. Other studies where the relativistic kinetic
theory has been used in the presence of curved space-times are refs. \cite{KD,KS}. However, this dependence
is small for non compact objects. We can list some values for the surface of some gravitational sources:
\begin{enumerate}
\item  Earth: $M_{\oplus}\approx 5.97\times
10^{24}$ kg; $R_\oplus\approx 6.38\times 10^6$ m;
${\vert\Phi(R_\oplus)\vert/c^2}\approx 7 \times 10^{-10}$;
\item   Sun: $M_{\odot}\approx 1.99\times
10^{30}$ kg; $R_\odot\approx 6.96\times 10^8$ m; ${\vert\Phi(R_\odot)\vert/c^2}\approx 2.2 \times 10^{-6}$;
\item   White dwarf: $M\approx 1.02 M_{\odot}$;
$R\approx 5.4\times 10^6$ m;
${\vert\Phi(R)\vert/c^2}\approx 2.8 \times 10^{-4}$;
\item   Neutron  star: $M\approx
M_{\odot}$; $R\approx 2\times 10^4$ m;
${\vert\Phi(R)\vert/c^2}\approx 7.5 \times 10^{-2}$.
\end{enumerate}

Lastly we comment the behavior of the transport coefficients with the help of the figures. We note that:

\begin{itemize}
  \item All transport coefficients (shear and bulk viscosities, thermal conductivity, diffusion and thermal-diffusion ratio) have larger values in the non-relativistic limiting case $\z_1\gg1$ (low temperatures) than those corresponding to the ultra-relativistic limiting case $\z_1\ll1$ (high temperatures).
  \item Shear and bulk viscosities and thermal conductivity  have smaller values in comparison with the corresponding ones for a single constituent. Setting particles of species 2 as the heavier ones, we note that the value of these transport coefficients decrease by increasing the concentration of species 2.
  \item The diffusion and thermal-diffusion coefficients increase when the concentration of the species with heavier particles grows.
  \item The thermal-diffusion rate is negative indicating that the constituent 1 tends to move into the warmer region and the constituent 2 towards the cooler region.
\end{itemize}

\section*{Acknowledgment}

G. M. K. acknowledges the Conselho Nacional de Desenvolvimento Cient\'{\i}fico e Tecnol\'ogico (CNPq, Brazil) and V. M.  the Consejo Nacional de Ciencia y Tecnolog\'{\i}a (CONACyT, M\'exico). The authors acknowledge to one of the referees for the careful reading of the manuscript and for the kindly observations.

\appendix

\section{Evaluation of the integrals}

In this appendix we show the main steps for the evaluation of the integrals listed in the appendix B. We start by introducing the definition of the total momentum $P^\mu$ and relative momentum $Q^\mu$ four-vectors through the relationships (see e.g. \cite{St})
 \ben\lb{A1a}
 P^\mu\equiv p_a^\mu+p_b^\mu,\qquad P^{\prime\mu}\equiv p^{\prime\mu}_a+p^{\prime\mu}_b,
 \qquad
 Q^\mu=p_a^\mu-p_b^\mu,\qquad
 Q^{\prime\mu}=p^{\prime\mu}_a-p^{\prime\mu}_b.
 \een
By using the energy-momentum conservation law i.e. $p_a^\alpha+p_b^\alpha=p'^\alpha_{a}+p'^\alpha_{b}$, above equations lead to
 \ben\lb{A2a}
P^\mu=P^{\prime\mu},\qquad
 P^\mu Q_\mu=(m_a^2-m_b^2)c^2,
 \qquad Q^2=P^2-{2(m_a^2+m_b^2)c^2},
 \een
 where the magnitudes of the total and relative momentum four-vectors are given by $P^2=P^\mu P_\mu$ and $Q^2=-Q^\mu Q_\mu$, respectively.
 From (\ref{A1a})  it follows the inverse transformations
  \ben\lb{A3a}
 p_a^\mu=\frac{P^\mu}{2}+\frac{Q^\mu}{2},\qquad
 p_b^\mu=\frac{P^\mu}{2}-\frac{Q^\mu}{2},
 \qquad
 p^{\prime\mu}_a=\frac{P^\mu}{2}+\frac{Q^{\prime\mu}}{2},\qquad
 p^{\prime\mu}_b=\frac{P^\mu}{2}-\frac{Q^{\prime\mu}}{2}.
 \een
The Jacobian of the transformation from $(p^\mu_a,p^\mu_b)$ to $(P^\mu, Q^\mu)$ is 1/8 so that $d^3p_a\, d^3p_b=d^3P\,d^3Q/8$.

If we introduce a space-like unit vector $\k^\mu$ which is orthogonal to $P^\mu$ ($\k^\mu P_\mu=0$)  the relative momentum four-vector can be decomposed as
 \ben\label{qmu}
Q^\mu=(m_a^2-m_b^2)c^2\,\frac{P^\mu}{P^2}
 +\frac{\k^\mu}{ P}\sqrt{P^4-2P^2(m_a^2+m_b^2)c^2+(m_a^2-m_b^2)^2c^4}.\qquad
\een
Here we shall restrict ourselves to the case where the rest masses of the particles of the constituents are not too disparate, that is $m_b\approx m_a$. So we have $(m_a^2+m_b^2)\approx 2m_am_b$ and terms of higher order are neglected. In this approximation $P^\mu Q_\mu=0,$ and the relative momentum four-vector (\ref{qmu}) and its modulus can be approximated by
 \ben\lb{A5}
 Q^\mu=Q\,{\k^\mu}, \qquad Q^2=P^2-4m_am_bc^2.
\een
This approximation transforms the invariant flux as
\ben\lb{A6}
F_{ba}=\sqrt{(p^\mu_ap_{b\mu})^2-m_a^2m_b^2c^4}=\frac{PQ}{2}.
\een

 Now the center-of-mass system is
 chosen where the spatial components of the total momentum
 four-vector vanish. Hence by introducing the representations $(P^\mu)=(P^0,{\bf 0})$
and $(Q^\a)=(0,{\bf Q})$  it follows the relationship
 \ben\lb{A7}
 F_{ba}\frac{d^3p_a}{ p_{a0}}\frac{d^3p_b}{ p_{b0}}=\frac{Q}{4\sqrt{g_0}}\frac{d^3P\,d^3Q}{P_0}.
 \een

 Furthermore, we write the element of solid angle  as  $d\Omega=\sin\Theta d\Theta d\Phi$, with $\Theta$ and
$\Phi$ denoting the polar angles of the spatial components of $Q^{\prime\alpha}$ with respect to
$Q^{\alpha}$ and such that $\Theta$ represents the scattering angle.
The differential cross sections $\sigma_{ab}$ depend on the scattering angle $\Theta$ and on the modulus of the relative momentum four-vector $Q$ so that  $\sigma_{ab}=\sigma_{ab}(Q,\Theta)$.
 In this work we are interested in analyzing a mixture where the differential cross sections are constant, which in the non-relativistic case corresponds to a mixture of hard spheres particles.
 Besides we may assume  without loss of generality that the spatial component of $Q^{\alpha}$ is in the direction of the axis $x^3$, so that we can represent $Q^{\alpha}$ and $Q'^{\alpha}$ as:
\ben\lb{A8}
(Q^{\mu})=Q\left(\begin{array}{c}0\\ 0\\ 0\\ 1
\end{array}\right),
\qquad
(Q^{\prime\mu})=Q\left(\begin{array}{c}0\\ \sin\Theta\cos\Phi\\
\sin\Theta\sin\Phi\\ \cos\Theta\end{array}\right).
\een

By taking into account the above premisses we can obtain the following results
\ben\lb{A9a}
\int(p_a^{\prime\mu}-p_a^\mu)d\Omega=\frac{1}{2}\int(Q^{\prime\mu}-Q^\mu)d\Omega=-2\pi\,Q^\mu,\quad
\\\lb{A9b}
\int(p_a^{\prime\mu}p_a^{\prime\nu}-p_a^\mu p_a^\nu)d\Omega=\frac{\pi}{3}Q^2\left(\frac{P^\mu P^\nu}{P^2}-g^{\mu\nu}-3\frac{Q^\mu Q^\nu}{Q^2}\right)-\pi(P^\mu Q^\nu+P^\nu Q^\mu).\quad
\een
Furthermore, it is also possible to perform the integrations in the spherical angles of
$Q^\mu$, denoted by $\theta$ and $\phi$. We write
\ben\lb{A10}
 (Q^{\mu})=Q\left(\begin{array}{c}0\\
\sin\theta\cos\phi\\  \sin\theta\sin\phi\\ \cos\theta\end{array}\right),
\een
with $d^3Q=Q^2\sin\theta\,d\theta\,d\phi\,dQ=Q^2d\Omega^\star dQ,$ where  $d\Omega^\star= \sin\theta d\theta d\phi$ denotes an element of
solid angle. From the above considerations it is easy to find the expressions for the integrals:
\ben\lb{A11a}
\int d\Omega^\star=4\pi,\; \int Q^{\mu} d\Omega^\star=\int Q^{\mu}Q^{\nu}Q^\sigma d\Omega^\star=0,
\qquad
\int Q^{\mu}Q^{\nu}d\Omega^\star={4\pi\over 3}Q^2
\left({P^{\mu}P^{\nu}\over P^2}-g^{\mu\nu}\right),
\\\lb{11b}
\int Q^{\mu}Q^{\nu}Q^\sigma Q^\tau d\Omega^\star=\frac{4\pi}{15}
Q^4\bigg[3\frac{P^\mu P^\nu P^\sigma P^\tau}{P^4}
-\frac1{P^2}\big(g^{\mu\nu}P^\sigma P^\tau+g^{\mu\sigma}P^\nu P^\tau+g^{\mu\tau}P^\sigma P^\nu
\\\no
+g^{\nu\sigma}P^\mu P^\tau+g^{\nu\tau}P^\sigma P^\mu+g^{\sigma\tau}P^\mu P^\nu\big)
+g^{\mu\nu}g^{\sigma\tau}+g^{\mu\sigma}g^{\nu\tau}+g^{\mu\tau}g^{\sigma\nu}\bigg].\qquad
\een

 The integration in the total momentum four-vector can be performed and for that end
 we consider a comoving frame where $(U^\mu)=(c/\sqrt{g_0},{\bf 0})$, a spherical
 coordinate system  and write  $d^3P=\vert{\bf P}\vert^2 d\vert{\bf P}\vert \sin\vartheta\, d\varphi\, d\vartheta$.
We introduce a new variable of integration
\ben\lb{A12}
x=\sqrt{\z_a\z_b\left(\frac{Q^2}{m_am_bc^2}+4\right)},\qquad Q\,dQ=\left(\frac{kT}{c}\right)^2x \,dx,
\een
where the range of this new variable is  $2\sqrt{\z_a\z_b}\leq x<\infty$. Hence, we can  write the time and spatial coordinates of the total momentum four-vector as
\ben\lb{A13}
P_0={kT\over c} \sqrt{g_0}\,xy, \qquad\vert{\bf P}\vert={kT\over c}\frac{x\sqrt{y^2-1}}{\sqrt{g_1}},
\qquad P=\frac{kT}{c}x.
\een
Hence, the element of integration becomes
\ben\lb{A14}
\sqrt{g_1^3g_0}\frac{d^3P}{P_0}=\left({kT\over c}\right)^2x^2\,\sqrt{y^2-1}\,\sin\vartheta\, d\varphi\, d\vartheta\,dy,\quad
\een
while the range of integration of the new variable $y$ is now $1\leq y<\infty$.

The integration  over the solid angle and over the variable $y$  can be performed and we obtain the following results:
\ben\lb{A15a}
\sqrt{g_1^3g_0}\int e^{-\frac{P^\mu U_\mu}{kT}}\frac{d^3P}{P_0}=4\pi\left({kT\over c}\right)^2xK_1(x),
\qquad
\sqrt{g_1^3g_0}\int e^{-\frac{P^\mu U_\mu}{kT}}\frac{P^\mu}{P^2}\frac{d^3P}{P_0}=4\pi{kT\over c}K_2(x)\frac{U^\mu}{c},
\\\lb{A15c}
 \sqrt{g_1^3g_0}\int e^{-\frac{P^\mu U_\mu}{kT}}\frac{P^\mu P^\nu}{P^2}\frac{d^3P}{P_0}=4\pi\left({kT\over c}\right)^2\left[xK_3(x)\frac{U^\mu U^\nu}{ c^2}-K_2(x)\,g^{\mu\nu}\right],
\\\no
 \sqrt{g_1^3g_0}\int e^{-\frac{P^\mu U_\mu}{kT}}\frac{P^\mu P^\nu P^\sigma}{P^2}\frac{d^3P}{P_0}=4\pi\left({kT\over c}\right)^3\left[x^2K_4(x)\frac{U^\mu U^\nu U^\sigma}{c^3}\right.
\\\left.
-xK_3(x)\left(g^{\mu\nu}\frac{U^\sigma}{c}
  +g^{\mu\sigma}\frac{U^\nu}{c}+g^{\nu\sigma}\frac{U^\mu}{c}\right)\right],
\\\no
\sqrt{g_1^3g_0}\int e^{-\frac{P^\mu U_\mu}{kT}}\frac{P^\mu P^\nu P^\sigma  P^\tau}{P^4}\frac{d^3P}{P_0}=4\pi\left({kT\over c}\right)^2\bigg[\frac{K_3(x)}{x}(g^{\mu\nu}g^{\sigma\tau}
+g^{\mu\sigma}g^{\nu\tau}+g^{\mu\tau}g^{\sigma\nu})
\\\no
-K_4(x)\bigg(g^{\mu\nu}\frac{U^\sigma U^\tau}{c^2}
+g^{\mu\sigma}\frac{U^\nu U^\tau}{c^2}+g^{\mu\tau}\frac{U^\nu U^\sigma}{c^2}+g^{\nu\sigma}\frac{U^\mu U^\tau}{c^2}+g^{\nu\tau}\frac{U^\mu U^\sigma}{c^2}
+g^{\sigma\tau}\frac{U^\mu U^\nu}{c^2}\bigg)
\\
+xK_5(x)\frac{U^\mu U^\nu U^\sigma U^\tau}{c^4}\bigg].
\een
Above $K_n(x)$ are  modified Bessel functions of second kind
\ben\label{bf}
K_n(x)=\left(\frac{x}2\right)^n\frac{\Gamma(1/2)}{\Gamma(n+1/2)}\int_1^\infty e^{-xy}\left(y^2-1\right)^{n-\frac12}dy,
\een
with $n=0,1,2,\dots$.

 Furthermore, in order to perform the integrations in the variable $x$  we need the following integrals of Bessel functions, where $\chi=2\sqrt{\z_a\z_b}$:
\ben\lb{A17a}
\int_\chi^\infty x^7 K_2(x)\,dx=\chi^5\left[\left(\chi^2+48\right)K_5(\chi)-4\chi K_6(\chi)\right],\quad
\int_\chi^\infty x^5 K_2(x)\,dx=\chi^4\left[\chi K_5(\chi)-6K_4(\chi)\right],
\\\lb{A17b}
\int_\chi^\infty x^3 K_2(x)\,dx=\chi^3K_3(\chi),\quad\int_\chi^\infty x K_2(x)\,dx=\chi K_1(\chi)+2K_0(\chi),
\quad
\int_\chi^\infty x^4 K_3(x)\,dx=\chi^4 K_4(\chi),
\\\lb{A17c}
\int_\chi^\infty x^8 K_3(x)\,dx=\chi^6\left[\left(\chi^2+8\right)K_6(\chi)-6\chi K_5(\chi)\right],\quad\int_\chi^\infty x^6 K_3(x)\,dx=\chi^5\left[\chi K_6(\chi)-8K_5(\chi)\right],
\\\lb{A17d}
\int_\chi^\infty x^2 K_3(x)\,dx=\chi^2 K_2(\chi)+4\chi K_1(\chi)+8K_0(\chi),
\quad
\int_\chi^\infty K_3(x)\,dx= K_2(\chi)+\frac2\chi K_1(\chi).
\een

\section{Expressions for the  elements of the matrices for binary mixtures}
From the above considerations we can perform the integrals
\ben\lb{A18}
\int \psi_b^\mu\mathcal{I}_{ab}\left[ \phi_a\right]\frac{d^{3}p_{a}}{p_{a0}}=\int \psi_bf_a^{(0)}f_b^{(0)}\left(\phi'_a-\phi_a\right)
 F_{ab}\sigma_{ab}d\Omega\sqrt{-g}
\frac{d^{3}p_{b}}{p_{b0}}\frac{d^{3}p_{a}}{p_{a0}},
\een
where $\phi_a=\phi_a(p_a^\mu)$, $\psi_a=\psi_a(p_a^\mu)$ are functions of the momentum four-vector of the particles $p^\mu_a$ and $f_a^{(0)}$ is the Maxwell-J\"uttner distribution function, which  in a comoving frame reads
\ben\lb{A19}
f_{a}^{(0)}=\frac{\n_{a}}{4\pi kTm_{a}^{2}c{K}_{2}\left(\zeta_{a}\right)}e^{-\frac{c\sqrt{m_{a}^{2}c^{2}+g_{1}\vert{\bf p}_{a}\vert^{2}}}{kT}}.
\een

In the following list the first identity is a general
integral and the second one is evaluated when similar masses of the species and hard-sphere
interaction are considered. The  resulting  expressions for the elements of the matrices for a binary mixture are:
\ben\lb{A20a}
\mathcal{A}_{12}=-\frac{c\Delta^{\mu\nu}}{3\n_{1}\n_{2}kT}\int p_{1\mu}\mathcal{I}_{12}\left[p_{2\nu}\right]\sqrt{-g}\frac{d^{3}p_{1}}{p_{10}}=\frac{16\pi kT \sigma_{12}}{3c   K_2(\z_1)K_2(\z_2)}h_{12}^I,
\\\lb{A20b}
\mathcal{A}_{11}=-\frac{c\Delta^{\mu\nu}}{3\n_{1}^{2}kT}\bigg[\sum_{b=1}^{2}\int p_{1\mu}\mathcal{I}_{1b}\left[p_{1\nu}\right]+\int p_{1\mu}\mathcal{I}_{11}\left[p_{1\nu}\right]\bigg]\sqrt{-g}\frac{d^{3}p_{1}}{p_{10}}
=-\frac{\n_2}{\n_1}\mathcal{A}_{12},\qquad \mathcal{A}_{22}=-\frac{\n_1}{\n_2}\mathcal{A}_{12},
\\\no
\mathcal{F}_{12}=-\frac{c\Delta^{\mu\nu}}{3\n_{1}\n_{2}kT^{2}}\int p_{1\mu}\mathcal{I}_{12}\left[\frac{\zeta_{2}}{\c_{p}^{2}}
\left(G_{2}-\frac{U_{\tau}p_{2}^{\tau}}{m_{2}c^{2}}\right)p_{2\nu}\right]
\sqrt{-g}\frac{d^{3}p_{1}}{p_{10}}
\\\lb{A20c}
=-\frac{16\pi k \sigma_{12}}{3c  K_2(\z_1)K_2(\z_2)\c_p^2}\left[2h_{12}^{II}-\z_2G_2h_{12}^I\right],
\\\no
\mathcal{F}_{11} = -\frac{c\Delta^{\mu\nu}}{3\n_{1}^{2}kT^{2}}\left\{ \sum_{b=1}^{2}\int p_{1\mu}\mathcal{I}_{1b}\left[\frac{\zeta_{1}}{\c_{p}^{1}}
\left(G_{1}-\frac{U_{\tau}p_{1}^{\tau}}{m_{1}c^{2}}\right)p_{1\nu}\right]\right.
\\\left.
+\int p_{1\mu}\mathcal{I}_{11}\left[\frac{\zeta_{1}}
  {\c_{p}^{1}}\left(G_{1}-\frac{U_{\tau}p_{1}^{\tau}}
 {m_{1}c^{2}}\right)p_{1\nu}\right]\right\}\sqrt{-g}\frac{d^{3}p_{1}}{p_{10}}=-\frac{\n_2}{\n_1}\mathcal{F}_{21},
\\\lb{A20d}
\mathcal{F}_{21}=-\frac{16\pi k \sigma_{12}}{3c  K_2(\z_1)K_2(\z_2)\c_p^1}\left[2h_{12}^{II}-\z_1G_1h_{12}^I\right],
\mathcal{F}_{22}=-\frac{\n_1}{\n_2}\mathcal{F}_{12},
 \\
\mathcal{H}_{12}=-\frac{c\Delta^{\mu\nu}}{3\n_{1}\n_{2}kT^{3}}
\int\frac{\zeta_{1}}{\c_{p}^{1}}\left(G_{1}-\frac{U_{\sigma}p_{1}^{\sigma}}
{m_{1}c^{2}}\right)p_{1\mu}\mathcal{I}_{12}\left[\frac{\zeta_{2}}{\c_{p}^{2}}\left(G_{2}
-\frac{U_{\epsilon}p_{2}^{\epsilon}}{m_{2}c^{2}}\right)p_{2\nu}\right]\sqrt{-g}
\frac{d^{3}p_{1}}{p_{10}}\no
\\\lb{A20e}
\qquad=\frac{16\pi k \sigma_{12}}{3cT K_2(\z_1)K_2(\z_2)\c_p^1 \c_p^2}
\left[G_1\z_1G_2\z_2h_{12}^I-2\left(G_1\z_1+G_2\z_2\right)h_{12}^{II}+2h_{12}^{III}\right],
\\\no
\mathcal{H}_{11} = -\frac{c\Delta^{\mu\nu}}{3\n_{1}^{2}kT^{3}}\Bigg\{ \sum_{b=1}^{2}\int\frac{\zeta_{1}}{\c_{p}^{1}}\left(G_{1}-\frac{U_{\sigma}p_{1}^{\sigma}}
{m_{1}c^{2}}\right)p_{1\mu}\mathcal{I}_{1b}\left[\frac{\zeta_{1}}{\c_{p}^{1}}
\left(G_{1}-\frac{U_{\epsilon}p_{1}^{\epsilon}}{m_{1}c^{2}}\right)p_{1\nu}\right]
\\
\qquad+\int\frac{\zeta_{1}}{\c_{p}^{1}}\left(G_{1}-\frac{U_{\sigma}p_{1}^{\sigma}}
 {m_{1}c^{2}}\right)p_{1\mu}
 \mathcal{I}_{11}\left[\frac{\zeta_{1}}{\c_{p}^{1}}
\left(G_{1}-\frac{U_{\epsilon}p_{1}^{\epsilon}}{m_{1}c^{2}}\right)p_{1\nu}\right]
\Bigg\} \sqrt{-g}\frac{d^{3}p_{1}}{p_{10}}
\no\\\lb{A20f}
\qquad=-\frac{16\pi k \sigma_{12}}{3cT K_2(\z_1)K_2(\z_2)(\c_p^1)^2 }\bigg\{\frac{\n_2}{\n_1}\left[(G_1\z_1)^2h_{12}^I-4G_1\z_1h_{12}^{II}+2h_{12}^{IV}\right]
+\frac{4\sigma_{11}K_2(\z_2)}{\sigma_{12} K_2(\z_1)}h_{11}\bigg\},
\\\lb{A20g}
\mathcal{H}_{22}=-\frac{16\pi k \sigma_{12}}{3cT K_2(\z_1)K_2(\z_2)(\c_p^2)^2 }\bigg\{\frac{\n_1}{\n_2}\left[(G_2\z_2)^2h_{12}^I-4G_2\z_2h_{12}^{II}+2h_{12}^{IV}\right]
+\frac{4\sigma_{22}K_2(\z_1)}{\sigma_{12} K_2(\z_2)}h_{22}\bigg\},
\\
\mathcal{R}_{12}=\frac{U_\mu U_\nu U_\sigma }{c^2\p_2}\int p_1^\mu p_1^\nu \mathcal{I}_{12}\Bigg[\frac{\partial\ln\z_2}{\partial \ln\c_v^2}\Bigg(\frac{ U_\tau p_2^\tau}{kT}
-\frac{3(\c_p^2+\h_2/T)}{\c_v^2}\Bigg)\frac{p_2^\sigma}{kT}
\Bigg]\sqrt{-g}\frac{d^{3}p_{1}}{p_{10}}\no
\\\lb{A20h}
\qquad=\frac{32\pi \p_1 \sigma_{12}}{c^2 K_2(\z_1)K_2(\z_2)}\frac{\partial\ln\z_2}{\partial\ln \c_v^2}\left[r_{12}^I-\frac{15(\c_p^2+\h_2/T)}{\c_v^2}r_{12}^{II}\right],\qquad
\\\no
\mathcal{R}_{11}=\frac{U_\mu U_\nu U_\sigma }{c^2\p_1}\Bigg\{\sum_{b=1}^2\int p_1^\mu p_1^\nu \mathcal{I}_{1b}\Bigg[\frac{\partial\ln\z_1}{\partial \ln\c_v^1}\Bigg(\frac{ U_\tau p_1^\tau}{kT}
-\frac{3(\c_p^1+\h_1/T)}{\c_v^1}\Bigg)\frac{p_1^\sigma}{kT}
\Bigg]
\\
\qquad+\int p_1^\mu p_1^\nu \mathcal{I}_{11}\Bigg[\frac{\partial\ln\z_1}{\partial \ln\c_v^1}\Bigg(\frac{ U_\tau p_1^\tau}{kT}-\frac{3(\c_p^1+\h_1/T)}{\c_v^1}\Bigg)\frac{p_1^\sigma}{kT}
\Bigg]\Bigg\}\sqrt{-g}\frac{d^{3}p_{1}}{p_{10}}\no
\\\lb{A20i}
\qquad=-\frac{64\pi \p_1 \sigma_{11}}{c^2 \z_1^2K_2(\z_1)^2}\frac{\partial\ln\z_1}{\partial\ln \c_v^1}r_{11}-\frac{32\pi \p_2 \sigma_{12}}{c^2 K_2(\z_1)K_2(\z_2)}\frac{\partial\ln\z_1}{\partial\ln \c_v^1}\left[r_{12}^{III}-\frac{15(\c_p^1+\h_1/T)}{\c_v^1}r_{12}^{II}\right],
\een
\ben\lb{A20j}
\mathcal{R}_{22}=-\frac{64\pi \p_2 \sigma_{22}}{c^2 \z_2^2K_2(\z_2)^2}\frac{\partial\ln\z_2}{\partial\ln \c_v^2}r_{22}-\frac{32\pi \p_1 \sigma_{12}}{c^2 K_2(\z_1)K_2(\z_2)}\frac{\partial\ln\z_2}{\partial\ln \c_v^2}\left[r_{12}^{III}-\frac{15(\c_p^2+\h_2/T)}{\c_v^2}r_{12}^{II}\right],
\\\lb{A20k}
\mathcal{K}_{12}=-\frac{c^3\Delta_{\mu\langle\sigma}\Delta_{\tau\rangle\nu}}{10\p_1 \h_1\p_2}
\int p_1^\sigma p_1^\tau\mathcal{I}_{12}\left[\frac{\z_2}{m_2\h_2}p_2^\mu p_2^\nu\right]
\sqrt{-g}\frac{d^{3}p_{1}}{p_{10}}=-\frac{64\pi c \sigma_{12}}{15kT G_1G_2K_2(\z_1)K_2(\z_2)}k_{12}^I,
\\
\mathcal{K}_{11}=-\frac{c^3\Delta_{\mu\langle\sigma}\Delta_{\tau\rangle\nu}}{10\p_1 \h_1\p_1}\Bigg\{
\sum_{b=1}^2\int p_1^\sigma p_1^\tau\mathcal{I}_{1b}\left[\frac{\z_1}{m_1\h_1}p_1^\mu p_1^\nu\right]
+\int p_1^\sigma p_1^\tau\mathcal{I}_{11}\left[\frac{\z_1}{m_1\h_1}p_1^\mu p_1^\nu\right]\Bigg\}\sqrt{-g}\frac{d^{3}p_{1}}{p_{10}}\no
\\\lb{A20l}
\qquad=\frac{64\pi c \sigma_{12}}{15kT \z_1^2 G_1^2K_2(\z_1)K_2(\z_2)}\frac{\n_2}{\n_1}k_{12}^{II}+\frac{64\pi c \sigma_{11}}{15kT K_3(\z_1)^2}k_{11},
\\\lb{A20m}
\mathcal{K}_{22}=\frac{64\pi c \sigma_{12}}{15kT \z_2^2 G_2^2K_2(\z_1)K_2(\z_2)}\frac{\n_1}{\n_2}k_{12}^{II}+\frac{64\pi c \sigma_{22}}{15kT K_3(\z_2)^2}k_{22}.
\een

Above we have introduced the following abbreviations with $\chi=2\sqrt{\z_1\z_2}$:
\ben
h_{12}^I=\frac{14}{\chi}K_3(\chi)
+\left(2+\frac{4}{\chi^2}\right)K_2(\chi),
\\
h_{12}^{II}=\left(\frac\chi2+\frac{34}{\chi}\right)K_3(\chi)
+\left(5+\frac{8}{\chi^2}\right)K_2(\chi),
\\
h_{12}^{III}=\left(\frac{9\chi}2+\frac{184}{\chi}\right)K_3(\chi)+
\left(28+\frac{\chi^2}4+\frac{32}{\chi^2}\right)K_2(\chi),
\\
f_{aa}=\left(\z_a+\frac{17}{\z_a}\right)K_3(2\z_a)+
\left(5+\frac{2}{\z_a^2}\right)K_2(2\z_a),
\\
h_{12}^{IV}=\left(\frac{9\chi}2+\frac{204}{\chi}\right)K_3(\chi)+
\left(30+\frac{\chi^2}4+\frac{48}{\chi^2}\right)K_2(\chi),
\\
h_{aa}=\frac5{\z_a}K_3(2\z_a)+\left(1+\frac2{\z_a^2}\right)K_2(2\z_a),
\\
k_{12}^I=\left(2+\frac{144}{\chi^2}\right)\frac{K_3(\chi)}{\chi}+
\left(20+\frac{64}{\chi^2}\right)\frac{K_2(\chi)}{\chi^2},
\\
k_{12}^{II}=\left(2\chi+\frac{134}{\chi}\right)K_3(\chi)+
\left(20+\frac{24}{\chi^2}\right)K_2(\chi),
\\
k_{aa}=\left(3+\frac{49}{\z_a^2}\right)\frac{K_3(2\z_a)}{\z_a}+
\left(15+\frac{2}{\z_a^2}\right)\frac{K_2(2\z_a)}{\z_a^2},\\
r_{12}^I=\left(\chi+\frac{96}{\chi}\right)K_3(\chi)+
\left(13+\frac{32}{\chi^2}\right)K_2(\chi),
\\
r_{12}^{II}=\frac{2K_3(\chi)}{\chi}+\frac1{5}
\left(1+\frac{8}{\chi^2}\right)K_2(\chi),
\\
r_{12}^{III}=\left(\chi+\frac{100}{\chi}\right)K_3(\chi)+
\left(13+\frac{48}{\chi^2}\right)K_2(\chi),
\\
r_{aa}=\z_aK_3(2\z_a)+2K_2(2\z_a),\qquad
\frac{\c_p^a+\h_a/T}{\c_v^a}=\frac{\z_a^2+6G_a\z_a-G_a^2\z_a^2}{\z_a^2+5G_a\z_a-G_a^2\z_a^2-1}.
\een


 \end{document}